# Global Predictions of the Universal Constant of Expansion

Charles B. Leffert, Emeritus Professor, Wayne State University
Contact: 1302 Wrenwood Drive, Troy, Michigan 48084

The new spatial condensation cosmological theory, based on the new universal constant of expansion $\kappa = G\rho t^2 = 3/32\pi$, is used to predict measurements of the supernova Ia data with no acceleration of the expansion rate of the universe. That goal was accomplished with the key assumption that radiation travels on great circles on the surface of the supporting 4-D ball towards the detector at the compounded velocity of $v_c = Hr - c$. The radiation has traveled great distances from emission to measurement in the expanding universe and is exactly suited to test the global predictions of a cosmological theory. Problems of the current big bang theory are discussed. It will be shown that a new method of analyzing the data does indeed show the simplicity and consistency of the new theory in contrast to the old Friedmann big bang model.

## 1.0 Introduction

The new spatial condensation (SC) theory of the expansion of the universe is being presented in two-papers. The first paper 1 [1] presented the derivation of the new universal constant, $\kappa = G\rho t^2 = 3/32\pi$, from which the entire theory flowed. It was shown that the SC-theory correctly predicts the cosmological parameters in agreement with those measured by the Wilkinson Microwave Anisotropy Probe WMAP [2] of age, $t_0$ =13.5 Gy; Hubble Constant, $H_0$ =68.6 km s$^{-1}$ Mpc$^{-1}$ and $\Omega_0$(mass) =0.28. However, the predictions of other key parameters were quite different: deceleration $q_0$ = 0.00842 and $\dot{R}$ = 1.005 c showing a near constant expansion rate of our universe with no present or future acceleration of the expansion.

Locally the effects of the expansion of our universe are negligible where Einstein's general relativity (GR) theory predictions have been confirmed. However the global vision of our universe is very different. The GR theory has a 4-D block time with only three spatial dimensions whereas the SC-theory has a 4-D space and separate asymmetric cosmic time. The GR-Friedmann equations have many solutions but the tight SC-theory has only one solution, so it is either correct or it is wrong. The heart of any cosmological theory is its scale factor as a function of time R(t) or t(R).

This paper is concerned with the global predictions R(t) of the new SC-theory for radiation that has traveled from its source over great distances in our expanding universe. The measurements of the radiation from a special class of distant exploding stars, called supernova Ia (SNIa) are just such measurements that are needed to test the new SC-theory.

In Section 2, the supernova Ia explosions will be briefly described. Section 3 will present some attempts by current cosmological models to predict those measurements. The transport of radiation through the 4-D space of the SC-theory will be developed in Section 4 and applied to current measurements of SNIa in Section 5, including a new method to analyze and compare the data to theory. Section 6 will test consistency of the new theory. Summary and conclusions are in Section 7. Table B1 on the derivation of the new transport equation is included as is Table A1, summarizing the equations of Paper 1.

## 2.0 Supernova Ia

Supernova Ia are a special class of exploding stars about which astrophysicists are confident that they know why and how they explode [3]. But it is the special manner in which they explode that produces approximately the same maximum absolute luminosity M that makes their measurements of value to the astronomer.

Before it explodes, a supernova star is a white dwarf star that has burned up its hydrogen and helium but does not have enough mass to ignite the beginning of nuclear burning of carbon towards iron. It also has a binary companion star that has been expanding and losing mass to the surface of white dwarf. When that added mass reaches a critical amount, the carbon reaction ignites and the white dwarf explodes.

A 2 ½ day rise and fall of the radiation pulse at emission can translate to a week long pulse at the detector and after further decay, the astronomer can then measure the redshift of the host galaxy. The maximum luminosity can be about equal to that of an entire host galaxy. Astronomers call these explosions "standard candles" indicating that when its radiation passes the Earth billions of years later, the measured pulse of radiation would give the distance to the source.

Distance is an elusive concept in an expanding universe. By the time the pulse of radiation reaches the astronomer's telescope, the exploded star has long ceased to exist and its companion has been expanded to a reception distance (RD) far greater than the emission distance (ED) to the world line of the Earth when the pulse was emitted. "World line" is used because even for emission at $Z = 1$, our Earth had not yet formed.

Astronomers have developed a magnitude scale [4] for the measurement of the intensity of received radiation and that radiation is received in a wide range of frequencies. Astronomers measure the "brightness" of objects by the radiant flux F of energy received at their detectors. At a distance of $r=1$ AU ($=1.496 \times 10^{13}$ cm) to the Earth, the measured flux of $F=1.360 \times 10^6$ erg s$^{-1}$ cm$^{-2}$ from Eq. (1) gives the Sun luminosity $L_{sun}=3.826 \times 10^{33}$ erg s$^{-1}$,

$$F = L/(4\pi r^2). \qquad (1)$$

But astronomers express their measurement in logarithmic units of magnitude "m". The Sun has apparent magnitude m=-26.81 mag. The absolute magnitude "M" of a star ($M_{sun} = 4.76$) is expressed as its apparent magnitude if placed at a distance of 10 parsec (pc) (1 pc = $3.0856 \times 10^{18}$ cm). Then the distance modulus "m-M" is,

$$(m - M) = 5\log (d/10 \text{ pc}). \qquad (2)$$

For the Sun at $d = r = 4.848 \times 10^{-6}$ pc, Eq. (2) gives $(m-M)_{sun} = -31.57$.

As the supernova Ia spherical pulse of radiation passes the astronomers telescope, astronomers apply a number of corrections to the decaying light curves of supernova Ia to estimate the peak apparent magnitude of these stellar explosions that are reported as a distance modulus (m-M) or as an effective magnitude,

$$m_{eff} = (m-M) + M. \qquad (3)$$

All supernova Ia exploding stars have about the same absolute magnitude M, which is more accurately measured on nearby explosions.

A 3-D radial pulse (few days width) of radiation of varying flux leaves the exploding SNIa source of luminosity L as an expanding 2-sphere. The arriving flux of radiation at a nearby radius r is $F = L/(4\pi r^2)$. If the pulse was emitted at time $t_{em}$ and arrived at time $t_0$, then radius $r = c(t_0 - t_{em})$. However, if the source was at an "emission distance, ED" at great redshift Z, then other effects, including the expansion of space and two relativistic effects, combine to reduce the energy flux. Let $r \to$ "$d_L$" be the "luminosity distance" increased for these other effects.

To build this expansion behavior into the prediction of the astronomer's apparent magnitude m, an effective luminosity distance dL is built into the flux equation, larger than 3-D r, to account for the received radiation.

Besides their estimated range of errors, the astronomers report two numbers for each SNIa: (1) the apparent effective magnitude $m_{eff}$ of the received radiation (or distance modulus m-M) and (2) the redshift Z of the source, often that of the host galaxy. For distant SNIa the measured flux increases in days and decays in weeks. The astronomers make corrections for some variations of the signal, which are not questioned or discussed here. The interested reader should consult the large literature [3].

The difference in topology between the new SC-theory (a 4-D spatially closed universe) and the BB-theory (a spatially flat 3-D universe) means that these additional effects of expansion will be derived differently. But apart from the details of these added effects, the evolution of the scale factor R with the expansion R(t), or t(R), must be correct.

From analysis of the new universal constant of expansion [1], the two theories differ in the number of spatial dimensions as well as a significant difference in the scaling with expansion of the third content of the universe. So it should be interesting to see how the difference in R(t) of the two theories effects their theoretical prediction of the SNIa $m_{eff}(Z)$.

Before developing the two theoretical SNIa predictions, we examine the SNIa data.

**2.1 Supernova Ia Data**

It is the measurements of the received radiation from the physical phenomena of the exploding stars called "supernova Ia (SNIa)" [3] that is used to test the global predictions of the SC-theory. Riess, et al.[5] presented a set of 185 SNIa measurements, which they say are of such good quality, they call it the "Gold and Silver" set of SNIa. In their Table **R**5 they present for each of their observed supernova; its name, its redshift Z, its distance modulus (m-M) and other information. In their Fig. (**R**4), the scatter of data points around their BB-predicted curve of m-M versus Z is more serious than it appears [6]. As more accurate SNIa measurements are made, recent analysis show they do not seem to fit any GR-cosmology [7].

A number of values of M have been reported for low-Z supernova Ia of which an SC-value M = -19.34 mag [8] was selected. The theoretical SC-linear curve of $m_{eff}$ versus Z is shown in Fig. 1 together with the reported SNIa data of Table **R**5 with $m_{eff}$ = (m-M) + (M = -19.34). The fit of the $m_{eff}$ data versus Z to the SC-theoretical curve for m is very similar to the fit of the BB-(m-M) data versus Z of the referenced Fig. **R**4 [5].

On the basis of equal fits of both theories to the raw data of this and other earlier SBIa data sets [9, 10], it follows that the former conclusion of acceleration of the expansion from the data was unjustified. This is because the conclusion depends upon

H(Z) and the SC-theory provides more than a satisfactory alternative with no acceleration.

Thus both theories will now be analyzed in more detail and a search made for a method of representing the data to demonstrate the better of the two theories.

**3.0 Transport of Radiation in the Big Bang Theory**

Carroll and Ostalie (C&O) [4] review big bang observational cosmology in their book *Modern Astrophysics*. In terms of redshift Z and scale factor R, radiation emitted at wavelength $\lambda_e$ at $R(t_e)$ is measured at wavelength $\lambda_0$ at $R(t_0)$.

$$R(t_0)/R(t_e) = R_0/R_e = \lambda_0/\lambda_e = (1 + Z), \tag{4}$$

This relation, although not $t_e/t_0$, is independent of the form of the scale factor $R(t)$. From the Robertson Walker (RW) metric for the consensus flat universe (k = 0), and neglecting lambda ($\Lambda=0$), gives the proper distance $d(t) = R(t) \varpi$ where $\varpi$ is the comoving coordinate (units of length), radially from the Earth ($\varpi=0$) and $d\theta = d\phi = 0$. For a photon emitted at $\varpi_e$ at time $t_e$, the metric ds = 0 and,

$$\int_{t0}^{te} \frac{cdt}{R(t)} = \int_0^{\varpi e} \frac{d\varpi}{\sqrt{1-k\varpi^2}}, \tag{5}$$

which leads for k = 0 to the reception distance RD for the source in units of $c/H_0$,

$$\frac{RD_{BB}}{c/H_0} = \frac{d(t_0)}{c/H_0} = \frac{R_0 \varpi}{c/H_0} = 2\left(1 - \frac{1}{\sqrt{1+Z}}\right), \tag{6}$$

and using Eq. (4), the big bang emission distance ED is,

$$\frac{ED_{BB}}{c/H_0} = \frac{d(t_e)}{c/H_0} = \frac{R(t_e)\varpi}{c/H_0} = \frac{2}{(1+Z)}\left(1 - \frac{1}{\sqrt{1+Z}}\right). \tag{7}$$

A pause is taken now to point out an important feature of Eq. (7), not mentioned by C&O that will also appear in the SC-theory by a different derivation. The first derivative of the right-hand side of Eq. (7) set to zero, reveals a maximum of $d(t_e)$ at $Z_m = 1.25$ [11]. This means that if Z is greater than $Z_m$, even though the photon is traveling radially towards Earth, it is really traveling radially away from Earth. During this backwards motion, the expansion takes a heavy tole in degrading the energy density of the SNIa pulse of radiation.

Further derivation by C&O yields for the comoving coordinate $\varpi(Z)$, in terms of the density parameter $\Omega$,

$$\frac{\varpi}{(c/H_0)} = \frac{2}{1+Z}\frac{1}{\Omega^2}\left[\Omega_0 Z - (2-\Omega_0)\left(\sqrt{1+\Omega_0 Z} - 1\right)\right]. \tag{8}$$

Equation (1) is adjusted by replacing r to define the theoretical luminosity distance by,

$$d_L^2 = L/4\pi F. \tag{9}$$

The relativistic effects are one factor of $(1 + Z)$ due to decrease in photon energy and another $(1 + Z)$ for the dilation of time interval between photons, so L is corrected by the factor $1/(1 + Z)^2$.

From the RW metric, $4\pi\varpi^2$ is the surface of a sphere at R = 1. Combining these factors, $F = L/4\pi\varpi^2(1 + Z)^2$ or $r \rightarrow d_L$ where,

$$d_L = \varpi(1 + Z), \tag{10}$$

to give finally for the luminosity distance in a matter dominated universe, where $\varpi$ is given either by Eq. (8) or its equivalent in terms of $q_0$ ($\Omega_0 = 2q_0$),

$$d_L = (2c/H_0)(1/\Omega_0^2)[\Omega_0 Z - (2 - \Omega_0)((1 + \Omega_0 Z)^{1/2} - 1)], \tag{11}$$

$$d_L = (c/H_0)(1/q_0^2)[q_0 Z - (1 - q_0)((1 + 2q_0 Z)^{1/2} - 1)]. \tag{12}$$

The SC-theory has $q \approx 0$ at the present and approaching zero into the future. Thus the incompatibility of the two theories is clearly evident from the second Eq. (12).

In the late nineties, with ordinary matter plus unknown dark matter equal to a total mass $\Omega_{mass} = 0.28$, Eqs. (11) or (12) did not fit the measured supernova Ia data at higher Z ($1 \leq Z \leq 1.5$), so astrophysicists added yet another unknown "lambda energy" $\Omega_\Lambda \approx 0.72$ to better fit the data and still maintain the popular k=0, $\Omega_0 = 1$, flat universe [9, 10]. The last $\Omega_\Lambda$-energy addition is said to produce a "negative pressure" and so sets our 3-D universe on a period of acceleration of the expansion rate.

This author's SC-theory had been published by this time [12, 13] without the need of these added energies, but two competing teams of astronomers [9, 10] had come to the same conclusion of accelerated expansion. Some scientists even now mention abandonment of the Friedmann equation [14].

**4.0 Transport of Radiation in the New SC-Theory**

After discovery of the universal constant of expansion kappa, $\kappa = G\rho t^2 = 3/32\pi$, in paper 1, the new SC-theory implied derivatives with respect to cosmic time t for such as the Hubble parameter H and the deceleration q. No integration of differential equations was required. We certainly know about radiation because our eyes and our instruments intercept its transport in epi-space and we have come to learn of its strange 3-D constant local speed c in the vacuum independent of the velocity of the 3-D source.

From our point of view radiation is confined to move in 3-D space. But it also moves in 4-D space, not because the vector $\vec{c}$ has any component towards the fourth dimension in 3-D space, but because our 3-D space itself is expanded in the radial direction $\vec{R}$. In the absence of a metric for the new SC-theory, a differential equation must be found for integration to account for all of these important motions. It turns out that the equation is fairly simple but its integration is not.

In the above big bang derivation of the luminosity distance $d_L$, the origin was placed at the exploding star and the expanding spherical pulse of radiation was followed out to the Earth to R = $R_0$ =1. For the SC-derivation of the luminosity distance $d_L$, the origin will be placed on the world line of the Earth and the emission distance ED is the radius r from that world line to the spherical pulse of photons. That distance of the advancing spherical pulse of radiation will be followed to that world line, of which a part terminates at the astronomer's detector. A sketch of the 4-D geometry should be helpful.

**4.1 Transport of Radiation in the 4-D Universe**

It is being proposed that our 3-D universe is the surface of an expanding 4-D ball. To view such a trajectory in two of the 4-D coordinates, imagine passing a plane through the 4-D ball that includes the axis of the telescope and the center of the 4-D ball to get the evolving view of Fig. 2. Each of the three circles represents only one of our three spatial dimensions at three different Z. The 4-D trajectory is a spiral that was started at Z=999, or $R/R_0 = 10^{-3}$, shortly after the decoupling of radiation and matter. All photons to an element of the telescope detector had been emitted to (or deflected onto) that trajectory. Therefore the spherical trajectory is also the trace of all emission distances SC-ED that deposit their photons at the detector on Earth at the present $t_0$.

At cosmic time t, the true 3-D distance (Mpc), ED, of the photons to the detector world line is measured on the arc of the circle at time t. Between the vertical world line of the detector (its atoms) and the spiral trajectory, the circle of maximum arc length is at $Z_m$ =1.7. As a second source, a star was placed on the Z = 1.7 circle where it crosses the spiral trajectory with emission distance ED. The world line of the star reaches the Z = 0 present circle at reception distance RD. For a star at rest in the expanding universe ($v_p$ = 0), always RD = (1 + Z) ED. The spiral is the 4-D light-distance to the source. Rotation of the 4-D cut as $R_0 \to R_0'$, $Z \to Z'$ for all present circles but still $Z_m'$ = 1.7 (see Fig. 12).

It is the uniform expansion of all 3-D space, through its expansion in the perpendicular fourth spatial dimension, that predominantly decreases the energy density of radiation in the SNIa pulse when $Z > Z_m$. Also, the 4-D expansion at the Planck level increases the wavelength of a photon, as well as, the distance between photons, both of which further decrease the energy in the SNIa pulse arriving to the detector.

It is the decrease in the Hubble parameter with increasing time that allows the trajectory of radiation to pass through the maximum distance and gain net negative velocity as r decreases and finally arrive at the detector at H = $H_0$ and r =0 with $v_c$ = c = 307 Mpc/Gy.

**4.2 SC-Equation for Transport of Radiation**

Radiation moves on great circles of the expanding 4-D ball of circumference C = $2\pi R$. Consider point p at rest in the CMB ($v_p$ = 0). Except for such as gravitational deflection, any radiation that will eventually pass through point p from direction $\vec{r}$ at distance r was moving toward p at compounded velocity,

$$v_c = dr/dt = Hr - c. \qquad (13)$$

The big bang distance ED of Eq. (7) can be determined by using the integrating factor $\alpha$ = $\exp[\int_{t_e}^{t} H(t')dt']$. As mentioned, the derivative with respect to Z of Eq.(8), set to zero, shows a maximum at Z = 1.25. A similar, but more complex expression for ED/(c/$H_0$) was derived for the SC-theory using the same integrating factor as shown in Table B.

That integration required the approximation of neglecting the radiation content of our universe at the present and future. The final expression also has a maximum $ED_m$ but at $Z_m = 1.7$ where $Hr = |c|$ as shown in Fig. 2.

The contribution of the radial expansion (c) to the SNIa radiation decay has two parts; (1) the normal $c(t_0 – t_{em})$ if there was no universe expansion and; (2) (RD-ED) = Z·ED if there was only 4-D universal expansion. The two relativistic effects are one factor of $(1 + Z)$ due to the decrease in photon energy and another $(1 + Z)$ for the dilation of time interval between photons, the same as for the BB-theory. Collecting terms (for later use with $t_{em}(Z)$ and $ED(Z)$ in units of $c/H_0$), gives for the luminosity distance and distance modulus, respectively, as shown in Table B,

$$dL = (c(t_0 - t_{em}) + Z·ED)(1 + Z), \tag{14}$$

$$m–M = 5 \log(dL/10 \text{ pc}). \tag{15}$$

The ratio of $dL/ED$ in Fig. 3 shows that an error $\delta Z$ in redshift Z at $Z \approx 2$ produces ten times the error in luminosity distance $\delta dL$ as the same $\delta Z$ error near $Z = 0$. The relative contribution and sum of the two terms of Eq. (14) are shown in Fig. 4.

It is not the current "distance" to the source of the radiation, but the distance to the oncoming radiation that is accounted for in the SC theory, because the expansion decreases the flux independent of whether the net motion $v_c = Hr\text{-}c$ is toward or away from the telescope. Using Eq. (13) in the SC-theory, such 3-D trajectories of the radiation are illustrated in Fig. 5 for radiation in our 3-D space propagating from the star at emission distance $ED(Z)$ at arbitrary $Z=10$ to the detector at $Z = 0$, starting at the present size of our universe $f = R/R_0 = 1$, but also starting at $f = 0.5$, $f = 2.0$ and $f = 3.0$. The first three trajectories are shown and the $f=3$ maximum distance (Mpc) from the detector ($v_c = 0$) was also at $Z_m = 1.7$.

With equal Z-intervals between the computer points, the number density of points in Fig. 5, indicates roughly the relative time spent before and after $v_c = 0$. With the SNIa radiation effectively stalled at $Z \geq 1.7$ in its approach to the detector, the expansion continues to degrade the energy density of the SNIa radiation and makes the predicted apparent magnitude very sensitive to small errors in the measured Z as will be seen.

Thus both theories take into account the backward motion of the photons at large emission distances, albeit with different functions of Z. The major difference is that of the evolution of the scale factor $R(t)$ due to the difference in topologies and contents and their scaling with the expansion.

Another interesting feature of the SC-theory is shown in Fig. 6. The SC-ED curve shows its maximum, $v_c = Hr\text{-}c = 0$ at $Z_m = 1.7$ as discussed above. But it is just at that value of $Z_m$ that the 4-D radius crosses the ED curve. In the future limit $\dot{R} \rightarrow c$ so $H = (\dot{R}/R) \rightarrow c/R$ or at $Z_m$, $v_c = (c/R)r = c$ or $r = R$. In words at $Z_m$, the 3-D distance to the world line of the detector equals the perpendicular 4-D radius of the universe.

Equation (14), in terms of Table B.1, and Eq. (15) were used to predict the SC-theoretical curve of Fig. 1. No parameters were adjusted to fit the data. But the scatter of the data is so large and the predictions of the theoretical curves are so similar that this type of plot cannot by itself distinguish the better theory.

Consider the shape of the curve of Fig. 1. For nearby SNIa one would expect more accurate measured values of both $m_{eff}$ and Z as shown. Even if the value of $m_{eff}$ was in small error, its vertical displacement would tend to keep it near the theoretical curve. On the other hand at high Z, the astronomer is still measuring the photon arrival rate with the same excellent equipment, but note that a small error in measurement of Z tends to shift the data point horizontally, which again minimizes the displacement from the theoretical curve.

Next consider the theoretical derivations. The final answers are essentially fixed by the final values of the important luminosity distance $d_L$. In both theories it is expressed in terms of Z and present values $H_0$ and $\Omega_0$, (or $q_0$) for the big bang theory, and $H_0$ and $\zeta_0$ for the SC-theory. The dependence on Z of the many factors that enter the derivation can tend to cancel in the final $d_L$, particularly if one is adding terms trying to force a fit to the data. Instead of the present value $H_0$, for which both theories tend to agree, one needs to find a way to represent the data in terms of H(Z) which would accentuate the large difference in the scale factor R(Z) between the two theories.

## 5.0 Representation in the $H/H_0$–$R/R_0$ Plane

For the BB-model and its problems of using the SNIa data to decipher the nature of its new dominant "dark energy," Padmanabhan and Choudhury [15] introduced the procedure of displaying the SNIa data in the dimensionless $\dot{a}$ - a plane [a = $R/(R_0=1)$, $\dot{a} = \dot{R}$]. The scatter of early SNIa data was so large that to show the predicted acceleration of the expansion they first "smoothed" the data and then showed the large scatter by error bars as reproduced in Fig.7. They show seven parametric curves for $\Omega_m$ for values 0 to 1. A very important observation for later contrast to the SC-theory is that the data do not follow any one of the seven big bang $\Omega_m$ theoretical curves for matter mass, but start at "a" = 1 along $\Omega_m$ =0.16, cross $\Omega_m$ =0.32, and the last point is greater than $\Omega_m$ = 0.48. There is only one actual $\Omega_m$, so this failure signals the first measured failure of general relativity. [16]

It also suggests a similar check of the new Gold and Silver data of Fig. 1 against the SC-theoretical curve in the dimensionless $H/H_0$–$R/R_0$ plane. For the Z-value reported for each SNIa, this meant calculating $R/R_0$ = 1/(1 + Z) and $H(Z)/H_0$ from Eq. (A.16) of Table A. Those new data and the SC-theoretical curve are shown in Fig. 8. Also shown are two SC-predicted curves for a hypothetical error in redshift of $\delta Z = \pm 0.1$. The bounded fit of data to SC-theory in Fig. 8, indicates a correct theory for a noisy signal with increasing range of noise at smaller R. The cause of the noise, or inherent increasing scatter of the data with increasing Z, is suggested in Section 5.1.

That the data in Fig. 8 **tend to follow** the SC-theoretical curve is very important in comparison to the BB-model of Fig. 7 with no trend to follow any one of the seven BB-theoretical curves. An important feature of this new plot of the data, is that it also shifts the data point differently depending on the theoretical model used, H(R). This is quite evident by noting the different vertical scales, and that both the theoretical curve and data of Fig. 8 would appear much higher if overlaid on Fig 7.

A recent "Supernova Legacy Survey (SLS)" of supernova Ia measurements has been reported [17]. These data were also analyzed in the $H/H_0$-$R/R_0$ plane and are shown in Fig. 9. They also exhibit the same general features as the "Gold and Silver" data.

## 5.1 Inherent Fluctuations

Stars twinkle on a characteristic scale of a fraction of a second, because of bending of light in its trajectory to our eyes through temperature fluctuations in our atmosphere. The spiral photon trajectory of Fig. 2 is a tortuous path through such as the gravitational fields of many massive objects whose distances of separation become ever less with smaller radius R of our universe. Also the sensitivity of measurement to small changes in Z become ever greater and to deflection of photons onto (and off of) that trajectory from greater distances at smaller R. Those massive objects of the past also slowly move, merge and grow. Thus collectively, astronomer's snapshot measurements are noisy because of the changing environment about that long photon trajectory.

**5.2 Consistency of the Data with the New Theory**

No claim is made that the following procedure can detect the source of errors.
If the SC-theory is correct, as strongly indicated in Fig. 8, it suggests that something might be gained by forcing the data to fit the theoretical model. There is no guide for how to move the points in the raw data of Fig. 1 to force them to be consistent with a model. The $H/H_0$ vs $R/R_0$ plot of Fig 8 does offer such a guide. Forcing a fit in the $H/H_0 - R/R_0$ plane of Fig. 8 would guarantee consistency with SC-H(Z) theory.
For the first variation of the Gold and Silver data, the measured $m_{eff}$ remains the same, and a fit of H to H(R(Z)) of Eq. (A.16) of SC-theory was forced in the H-R plane of Fig. 8. Because of the sensitivity of $m_{eff}$ to Z, an iterative calculation was used to find the new value of Z' such that its $(H/H_0)'$ was within a given tolerance $\vartheta$ of the SC-theoretical value (curve of Fig. 8). All of the data points did converge to the theoretical curve within the tolerance of $\vartheta = 0.005$.
The 185 values of the new $\Delta Z = Z'-Z$ were saved. The average deviation from the theoretical curve was $<\Delta Z> = -0.0027$. The $\Delta Z$ values were then applied to the $m_{eff}$ data of Fig. 1 and the final plot of $m_{eff}$ versus the corrected Z' shows in Fig. 10 all of the points on the theoretical curve of $m_{eff}$ vs Z. Now the Hubble parameter value H(Z) for each point agrees with theory which was not the case for Fig.1. A fit was also forced by variation of $m_{eff}$ holding Z constant, and likewise, an excellent fit of all of the data points was obtained.
Such manipulation of the raw data cannot determine the source of errors in the data, but it does suggest the theory is correct. The forced fit of data to theory accomplished in Fig. 10, is not likely to be accomplished in the big-bang model with the theoretical uncertainty shown in Fig.7 since none of the seven models shows a fit to the data.

**6.0 Theoretical Consistency of the New Theory for Projected Measurements**

The radiation of past supernova Ia, that our astronomers have measured, is still traversing through the universe. In principle, astronomers of the future, and some of the past, could measure the radiation of the same supernova Ia as it reaches other planets.
If theory has correctly captured the key variation of the scale factor with cosmic time R(t), then one should be able to project the origin for measurements $R_0$ to any new past or future $R_0'$ and correctly predict measurements of passing supernova radiation at any other R(t). The adjective "passing" excludes R(t) before the supernova occurred and radiation that has not yet arrived. The word "measurement" implies an un-lensed signal strength to meet present instrument limitations of about $m_{eff} < 26$ mag.
To orient the reader for the calculations to follow, consider the SC – R(t) plot of Fig. 11. Both the low-Z, 18 point, Hamuy et al, SNIa data and the higher-Z, 42 point

Perlmutter et al, data were taken from Tables 2 and 1, respectively, of the on-line LANL 41081 preprint [18], (later published [9]). To illustrate for later exercises, just the low-Z Hamuy data will be used here.

The Hamuy data were taken over a 3-year period (1990-1993) [19] and indicates the radiation had been traveling for an average of about one billion years. So only from a planet in our Galaxy could all 18 of these SNIa data have been measured. Our Galaxy has a diameter of about 30 kpc and it takes radiation about 100,000 years to cross so, in principle, other Galactic hypothetical astronomers could measure all 18 of these SNIa. Since these SNIa came to Hamuy's team from many directions, it is not likely they would be measured in the same order.

Of the 18 SNIa, SN1992bh was selected with $Z = 0.045$ and $m_B^{corr} = 17.61$. For that Z and the present universe of $t_0 = 13.5$ Gy, the SC-equations predict $dL/(c/H_0) = 0.0470$ and $m_{eff} = 17.56$ at $t_{em} = 12.89$ Gy and $H_{em} = 71.73$ km s$^{-1}$ Mpc$^{-1}$.

Next consider in Fig. 11 the look-forward time of 10 billion years at $t_0' = 13.5 + 10 = 23.5$ Gy. A quick calculation for $R_0'/R_0 = 1.70 = f$, gave $t_0' = 23.49$ Gy on the theoretical curve $R(t)$ and $Z' = -0.4118$ in our future with $H_0' = 40.24$ km s$^{-1}$ Mpc$^{-1}$.

Figure 12 in two of the 4-D dimensions was constructed to show how past 4-D radiation trajectories appear with the expansion and how values of the parameters change as our universe expands. The left spiral shows the parts of our universe that astronomer would observe if the telescope was pointed in exactly the opposite direction. Since we are following the same group of SN1992bh photons sampled on Earth, the same cut through the 4-D ball is just rotated about 24 degrees clockwise to put the future observer at the top. Only $Z' = 1.7$ maintains its relative position with $R_0'$ but all parameter values change with the new $R_0'$.

The densities at the new $R_0'$ were scaled according to Table A. The new much larger $Z'$ of SN1992bh is simply $Z' = f(1 + z) – 1 = 0.776$ and all information is now known at new origin $R_0'$ to calculate the details at SN1992bh from our much larger universe. Those details are: $dL/(c/H_0) = 1.221$, $m_{eff} = 25.79$, $t_{em} = 12.88$ Gy and $H_{em} = 71.74$ km s$^{-1}$ Mpc$^{-1}$. The last two values are in good agreement with those calculated from our present universe $R_0$. The new values of $dL/(c/H_0)$ and $m_{eff}$ and similar values for the other 17 SNIa will be mentioned later.

Next consider the changing general features of SNIa radiation as it expands radially in the expanding universe. Spatial condensation of the surface of the 4-D ball maintains a spatially uniform and isotropic 3-D universe and thus Z represents the size R of our universe independent of the direction of internal 3-D motion. If one has correctly modeled all of the factors that decrease the energy flux of the radiation as it expands, then the expression for the luminosity distance $dL/(c/H_0)$ (and distance modulus m-M) of that radiation as a function of Z, should appear, respectively, on the same rising theoretical curves.

Now consider all 18 of the Hamuy SNIa. The theoretical universal curve was calculated for the luminosity distance $dL/(c/H_0)$ and is shown in Fig. 13. Also shown are the measurements of Hamuy, et al. projected to future $R_0'$ origins where the residuals have been carried along with the projected present theoretical $m_{eff}$ values. The projected $dL/(c/H_0)$ data to $f = R_0'/R_0 = 1, 2, 3$ and 6 show that a specific SNIa point just moves up this curve ($R_0'$ dependent) as the universe expands.

Of course, future measured values of flux decrease, (effective magnitude $m_{eff}'$ increases) for the same radiation. These higher projections far exceed the present instrumentation limitation of $m_{eff}(max) = 26$ mag for SNIa.

The above exercise predicted measurements of existing SNIa radiation for future astronomers. One could also ask what the SC-theory predicts for existing SNIa radiation by possible astronomers of the past. For this exercise, seven "gold, high Z" (1.140 to 1.551) SNIa measurements [5] were selected. In Fig. 1 they were, counting from the right, SNIa 2, 3, 4, 5, 6, 7 and 9 as shown in Fig. 14. Similar calculations were made to predict measurements for $f = R/R_0 = 0.5$. The past predicted values of $dL/(c/H_0)$, as shown in Fig. 14, were on the same $dL/(c/H_0)$ curve as in the Fig. 13. As a final test calculation, f was decreased to 0.40 (t = 5.0 Gy) corresponding to $Z = 1.5$ and, indeed, only the highest SNIa of $Z = 1.551$ survived the calculation. The six other lower-Z SNIa were rejected because at $Z = 1.5$ the "past astronomers" were in our universe before the six lower-Z SNIa occurred.

**7.0 Summary and Conclusions**

In short, the Friedmann equations do not globally fit our 3-D universe. The new universal constant equations of the SC-theory **do fit** our 3-D universe, globally!

Locally, the effects of the expansion are not measurable and do not challenge the local GR theory. The only local effect that does challenge the GR theory is the predicted singularity inside a black hole, which is not present in the SC-theory.

The 4-D spatial version of our 3-D universe as the surface of an expanding 4-D ball was shown to be consistent geometrically with the universal constant of expansion. It also produced a successful 4-D theory and predicted acceptable measurements for the transport of radiation from distant supernova Ia to our astronomer's detectors.

The replacement of dark matter with x-stuff or dark mass that scales as $R^{-2}$ together with a new definition of cosmic time in paper 1 led to a simple constant expansion rate (rather than the GR acceleration). In turn in this paper, those changes led to universal curves for the luminosity distance dL and distance modulus m-M where the supernova Ia radiation is predicted to move steadily along these curves as our universe expands.

Consistency was shown where the measurements of a radial expanding pulse of a supernova Ia by our astronomers could be predicted for some measurements of earlier or later hypothetical astronomers in other parts of our universe. From their new coordinates ($R_0' \rightarrow R_0$), they can predict correctly the measurements of our (to them hypothetical) astronomers given only our $R_0'$, $Z_{em}'$ of the SNIa and the SC-theory.

No cosmological constant, dark matter, dark energy and no acceleration of the expansion rate were needed in the new theory. All that was needed were a correct reading of the universal constant of expansion kappa, $\kappa = G\rho t^2 = 3/32\pi$, and the 3-D radial transport equation, $v_c = Hr \pm c$, for radiation in the surface of an expanding 4-D ball.

**Acknowledgement**

The author thanks his good friend, Emeritus Professor Robert A. Piccirelli, for editorial comments.

# Table A: Summary of the SC-Cosmological Theory

The scale factor R has units of length for our 3-sphere, spatially 3-dimensional expanding universe; G is the gravitational constant; c is the local speed of light; and H is the Hubble parameter. Present values have subscript 0 and cgs units are assumed. Other subscripts include: r=radiation, m=matter and x=dark mass (not dark matter). Pertinent equations of the new theory [hereafter: "SC-theory"] are listed in Table A1.

### Table A.1  Derivation of Theory

| | | |
|---|---|---|
| Universal constant: | $\kappa = Gt^2\rho = Gt_0^2\rho_0 = 3/32\pi$. | (A.1) |
| From $T_0$=2.726 K: | $\rho_{r0} = 9.40 \times 10^{-34}$ g cm$^{-3}$. | (A.2) |
| From nucleosynthesis: | $\rho_{m0} = 2.72 \times 10^{-31}$ g cm$^{-3}$. | (A.3) |
| Present age: | $t_0 = 13.5$ Gy $= 4.260 \times 10^{17}$ s. | (A.4) |
| From (A.1): | $\rho_0 = (\kappa/G)/t_0^2$ g cm$^{-3}$. | (A.5) |
| From (A.6): | $\rho_{x0} = \rho_0 - \rho_{r0} - \rho_{m0}$ g cm$^{-3}$. | (A.6) |
| Present scale factor: | $R_0 = ct_0(\rho/\rho_{x0})^{1/2}$ cm. | (A.7) |
| Redshift Z (Input): | $R = R_0/(1+Z)$ cm. | (A.8) |
| Radiation density: | $\rho_r = \rho_{r0}(R_0/R)^4 = \rho_{r0}(1+Z)^4$. | (A.9) |
| Matter density: | $\rho_m = \rho_{m0}(R_0/R)^3 = \rho_{m0}(1+Z)^3$. | (A.10) |
| Dark Mass density: | $\rho_x = \rho_{x0}(R_0/R)^2 = \rho_{x0}(1+Z)^2$. | (A.11) |
| Total density: | $\rho(R) = \rho_r + \rho_m + \rho_x$. | (A.12) |
| Cosmic time: | $t(R) = + (t_0^2 \rho_0/\rho(R))^{1/2}$. | (A.13) |
| From time derivative: | $\rho_2 = 2\rho_r + 3/2\ \rho_m + \rho_x$. | (A.14) |
| From time derivative: | $\rho_3 = 4\rho_r + 9/4\ \rho_m + \rho_x$. | (A.15) |
| From time derivative: | $H = \dot{R}/R = (\rho/\rho_2)/t$. | (A.16) |
| Expansion Rate: | $\dot{R}/c = (R/ct)(\rho/\rho_2)$. | (A.17) |
| Deceleration, $q = -\ddot{R}R/\dot{R}^2 =$ | $(1/Ht)(-1 + [3 - 2(\rho\rho_3/\rho_2^2)])$. | (A.18) |

The scaling with the expansion of radiation, Eq. (A.9), and matter, Eq. (A.10), are borrowed from the big bang model, $\kappa$ has the same value for early Friedmann radiation.

The postulated scaling, Eq. (A.11), of the new and now dominant stuff called "dark mass," is the key signature of this new cosmological theory. Its density decreases with the expansion but its total mass, always in individual clumps, increases with the expansion. It is not a 3-D substance and so does not interact with radiation or matter except gravitationally, where it certainly contributes to the local curvature of 3-D space. The distribution of these miniscule dark mass seeds at the beginning of the expansion sets the pattern for the present large-scale structure, including voids, and contributes to the early formation of black holes and fit to supernova Ia data for $t_0$=13.5 Gy with no acceleration of the expansion rate.

The basic postulate for cosmic time, Eq. (A.13), was made in terms of partial times $\Gamma_i$ where $t^{-2} = \sum_i \Gamma_i^{-2}$ and $\Gamma_i^2 = (\kappa/G)/\rho_i(Z)$ where $\rho_i$ are given by Eqs. (A.9) to (A.11). With age set to $t_0$=13.5 Gy, the SC-theory predicted the following values for the present cosmological parameters: $R_0$=1.354x10$^{28}$ cm, $H_0$=68.6 km s$^{-1}$ Mpc$^{-1}$, $\Omega_B$=0.031, $\Omega_{DM}$=0.248, $\Omega_{DM}/\Omega_B$=8.0, $(\dot{R}/C)$=1.005 and $q_0$=0.0084 (i.e., approaching steady-state expansion), all within the range of uncertainty of our astronomer's measurements.

# Table B: 4-D Transport of Radiation in the SC-Theory

**Goal**: In terms of redshift Z, find a general expression for the 3-D emission distance ED from the radiation source to the world line of the detector and in terms of ED, a 4-D expression of the luminosity distance $d_L$. First, follow the photons and integrate $v_c dt = (Hr - c)dt$ from emission time $t_e$ at $Z_e$ at the source to time t at Z. Since $X_4 = 0$ at $Z = 0$, get positive $X_4$ by difference.

**Approximation**: For Z near zero and into the future, ignore low radiation density, $\rho_{r0} = 0$, and set $\varsigma_o = \rho_{x0}/\rho_{m0}$. So from Eq. (A.16) of Table A, $t_0 H_0 = 2/3 (1 + \varsigma_o)/(1 + 2/3\, \varsigma_o)$. For the integrating factor use,

$$\alpha(t) = \exp\left[\int_{t_e}^{t} H(t')dt'\right]. \tag{B.1}$$

From time derivatives of Eq. (A.13) and Eq. (A.12) of Table A, $Hdt = -dZ/(1 + Z)$, so,

$$\alpha = \exp[\ln[(1 + Z_e)/(1 + Z)) = (1 + Z_e)/(1 + Z). \tag{B.2}$$

And again from time derivative of Eq. (A.13) of Table A,

$$dt/t_0 = -((3/2\,(1 + \varsigma_o)^{1/2}/(1 + Z)^2[(1 + Z) + 2/3\,\varsigma_o]/[(1 + Z) + \varsigma_o]^{3/2})\,dZ. \tag{B.3}$$

$$\text{Let } X_4(Z) = ct_0 \int_{t_e}^{t} dt'/t_0\, \alpha(t')^{-1} = c\int_{Z_e}^{Z} \left(dt/t_0\right)\left((1+Z)/(1+Z_e)\right) dZ, \tag{B.4}$$

$$\int \frac{dx}{x(a+bx)^{m/2}} = 1/a \int \frac{dx}{x(a+bx)^{(m-2)/2}} - b/a \int \frac{dx}{(a+b)^{m/2}},$$

$$\int \frac{dx}{x(a+bx)} = \frac{1}{\sqrt{a}} \ln\left(\frac{\sqrt{a+bx}-\sqrt{a}}{\sqrt{a+bx}+\sqrt{a}}\right), \qquad \int \frac{dx}{(a+bx)^{3/2}} = \frac{-2}{(a+bx)^{1/2}}.$$

Integration gives $X_4(Z)$ in terms of $X_4(Z_e)$. But $X_4(Z=0) = 0$, so one can solve for $X_4(Z_e)$ and substituting back gives the desired 4-D distance to any source $X_4(Z)$.

$X(Z) = (c/H_0)(2/3[(1+\varsigma_o)/(1+2/3\,\varsigma_o)]\, ED$,

$$\text{and}\quad ED = [(1 + \varsigma_o)^{1/2}/(1+Z)][AZ - A0], \tag{B.5}$$

$$\text{where}\quad AZ = \left(\frac{1}{\sqrt{\varsigma_0}}\right) \ln\left(\frac{\sqrt{(\varsigma_0 + (1+Z))} - \sqrt{\varsigma_0}}{\sqrt{(\varsigma_0 + (1+Z))} + \sqrt{\varsigma_0}}\right) - \frac{1}{\sqrt{(\varsigma_0 + (1+Z))}}, \tag{B.6}$$

$$A0 = \left(\frac{1}{\sqrt{\varsigma_0}}\right) \ln\left(\frac{\sqrt{(\varsigma_0 + 1)} - \sqrt{\varsigma_0}}{\sqrt{(\varsigma_0 + 1)} + \sqrt{\varsigma_0}}\right) - \frac{1}{\sqrt{(\varsigma_0 + 1)}}. \tag{B.7}$$

From the text, in units of $c/H_0$, $d_L = (c(t_0 - t_e) + Z\, ED)(1 + Z)$,
$c(t_0 - t) = ct_0(1 - t/t_0)$ and $t/t_0 = 2/3(1 + \varsigma_o)/(1 + 2/3\,\varsigma_o)$,
so, $c(t_0 - t) = (c/H_0)[2/3(1 + \varsigma_o)/(1 + 2/3\,\varsigma_o)$ or, \hfill (B.8)

$$\frac{d_L}{(c/H_0)} = \left(2/3\right)\left((1+\varsigma_0)/(1+2/3\,\varsigma_0)\right)\left[\left(\frac{1}{((1+Z)+\varsigma_0)^{1/2}}\right) + ZED\right](1+Z). \tag{B.9}$$

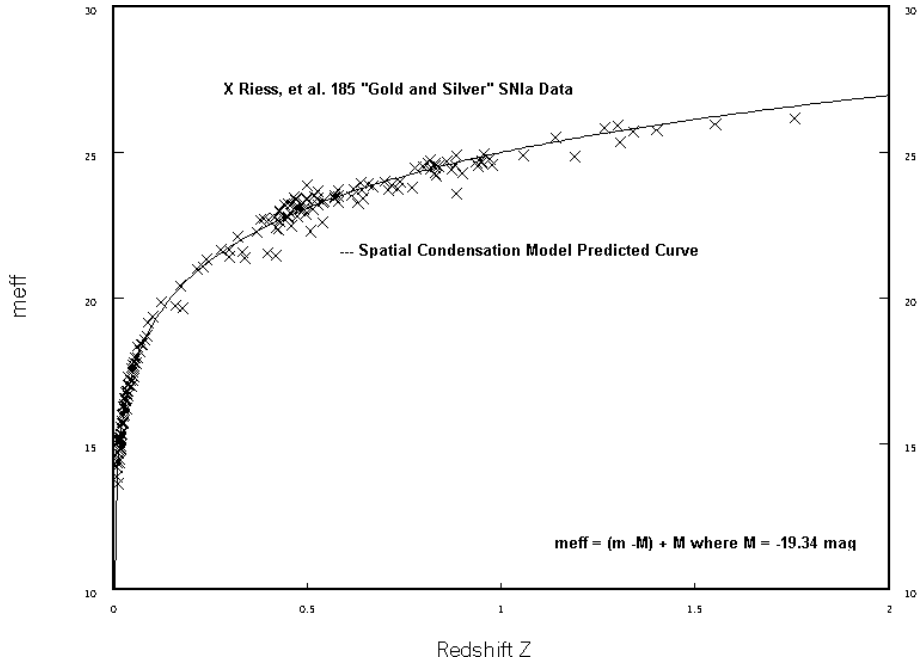

Fig. 1 A set of published 185 "gold and silver" values of distance modulus (m-M) of Riess et al, [5] were selected and presented here as m-M plus the constant absolute magnitude of the source, M = -19.34. The scatter of the data is still too excessive to provide information on the unknown "dark energy" used by the BB–model.

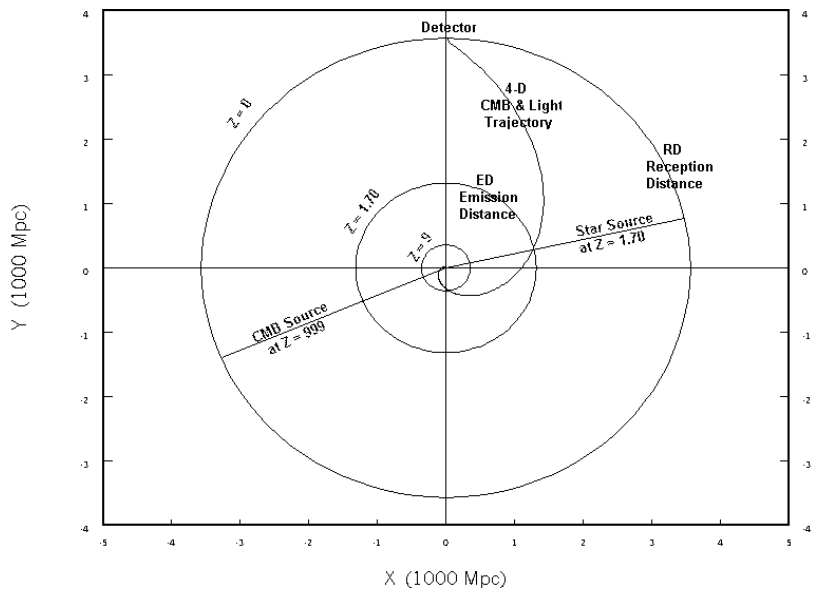

Fig. 2 This slice through the expanding 4-D ball, containing its center and the axis of the telescope pointing towards the incoming SNIa radiation, shows the 4-D spiral of the trajectory of the measured photons. Radiation must travel on an outer great circle of the 4-D ball but the circle is expanding perpendicular to the photon motion. The net effect of both motions is the spiral trajectory of the photons through 4-D space.

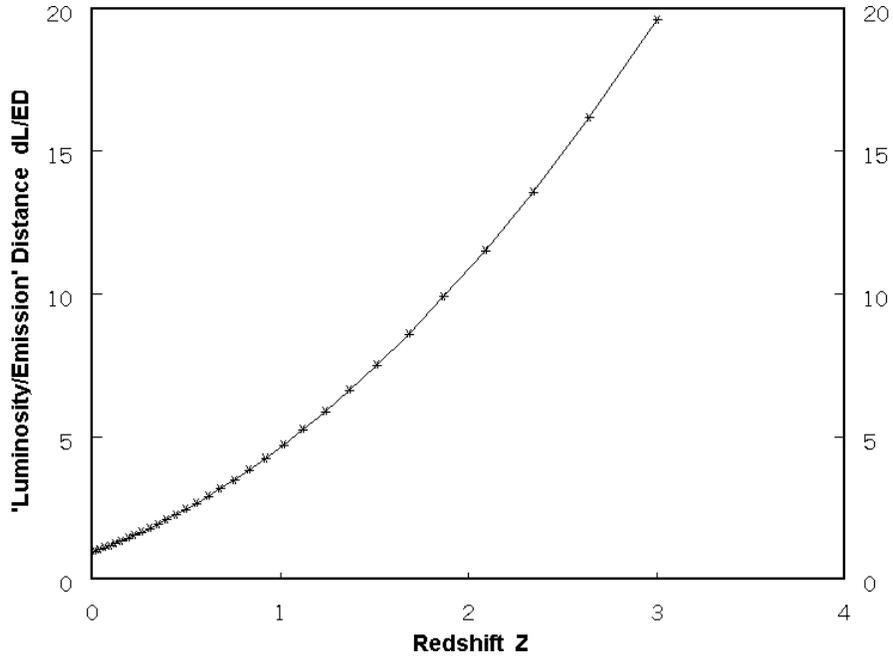

Fig. 3 For a nearby source of radiation, the flux at distance r is $F = L/4\pi r^2$ where L is the luminosity of the source. The luminosity distance is a theoretically corrected r to a larger value $d_L$ so that $F = L/4\pi d_L^2$ predicts the measured flux F. The SC-theoretical emission distance is ED at Z and the curve shows the rapid growth of $d_L$/ED with increasing Z.

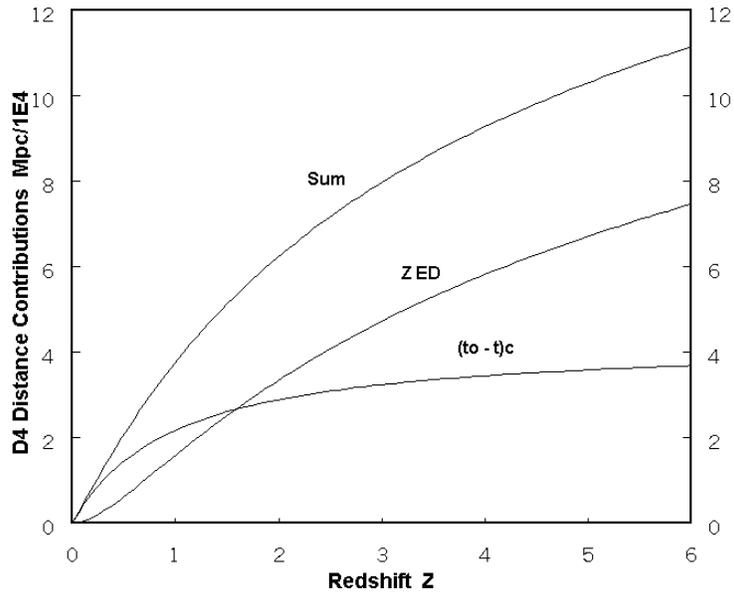

Fig:4 Radiation decay due to expansion overtakes that due to photon velocity. Apart from the relativistic effects that increase $d_L$ (see text), these curves show the SC-contributions of the speed of light $(t_0 - t)c$ and the expansion Z ED to the total distance of the spiral trajectory through 4-D space.

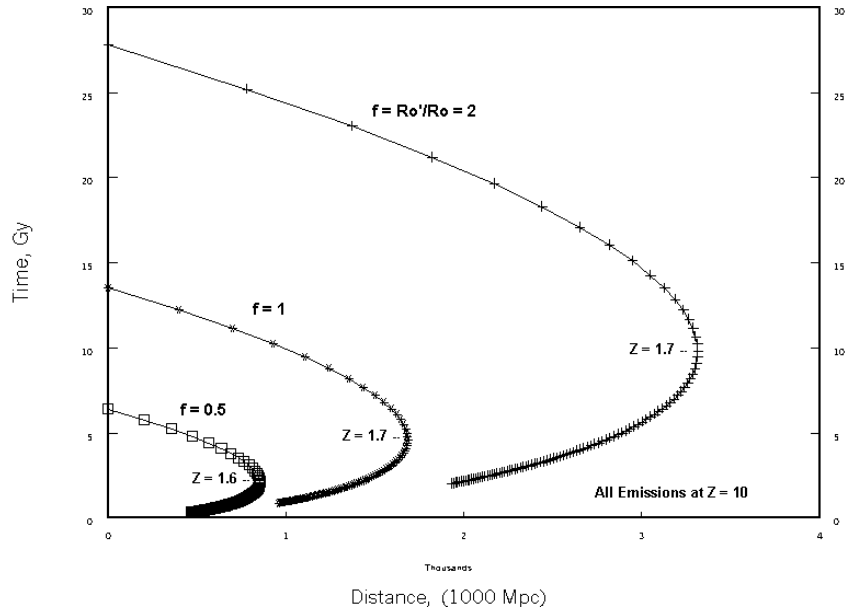

Fig. 5 In the SC-equations, the simple compounded velocity of light $v_c = Hr - c$ in 3-D space produces a maximum value of r at $Z = 1.7$, independent of the size of the universe into the future. The density of points for $Z > Z_m$ shows great loss of energy density due to the expansion while the radiation makes no net advance toward the detector.

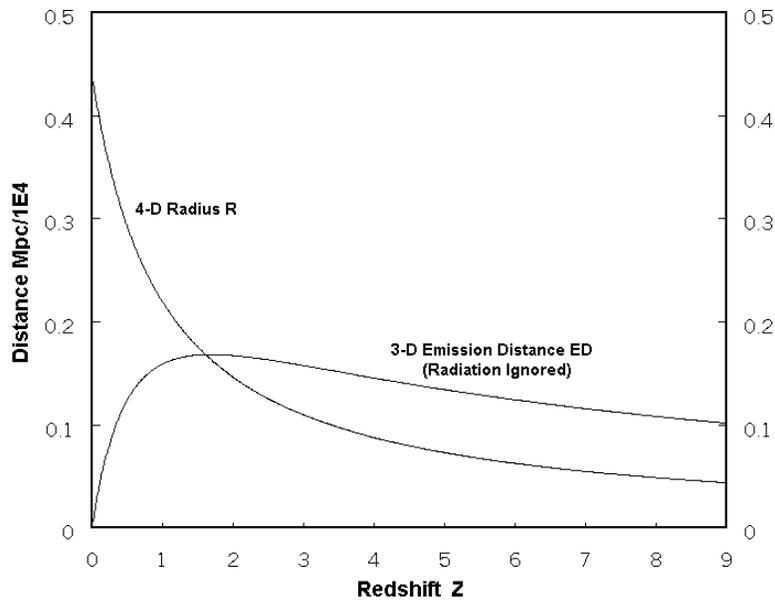

Fig. 6 At emission $t_{em}$, ED is the arc distance to the past world line of the detector. For emission at Z less than $Z_m = 1.7$, ED increases with increasing Z but the photon gains net velocity toward the detector. For $Z > Z_m$, as the photon travels at local velocity c towards the detector, the expansion actually carries it further away from the detector. Now, and surprisingly into the future, the radius of the expanding 4-D ball R equals ED at $Z_m$.

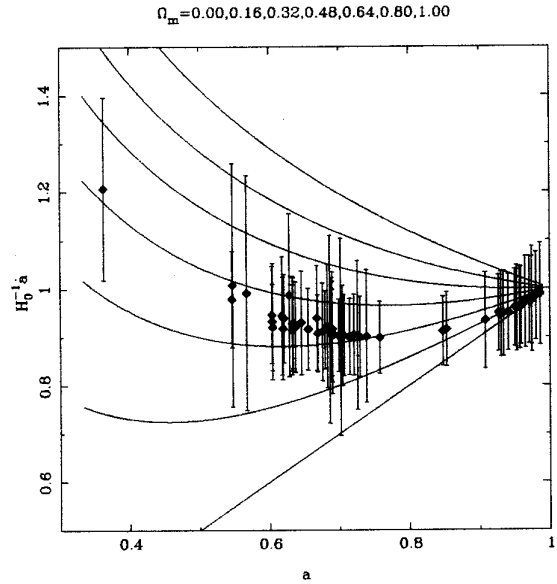

**Figure 3.** The observed supernova data points in the $\dot{a} - a$ plane for flat models. The procedure for obtaining the data points and the corresponding error-bars are described in the text. The solid curves, from bottom to top, are for flat cosmological models with $\Omega_m = 0.00, 0.16, 0.32, 0.48, 0.64, 0.80, 1.00$ respectively.

This Fig. 3 of astro-ph/0212573 was reproduced with the kind permission of author T. R. Choudhury.

Fig. 7 The two important points are: (1) the data do not to follow any one of the seven big-bang theoretical curves for $\Omega_m$; and (2) this new procedure of plotting in the $\dot{a}$ - a plane is most useful in analyzing the SC-equations in the equivalent $H/H_0 – R/R_0$ plane.

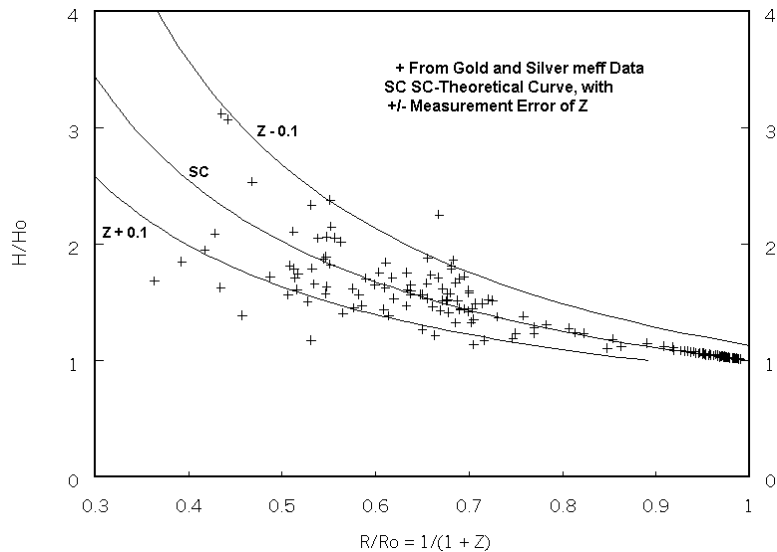

Fig. 8 For the Z-value of each SNIa, $R(Z)/R_0$ and $H(Z)/H_0$ are calculated and displayed in this new dimensionless plot in comparison with the SC-theoretical curve. Much more scatter of the data is shown than in Fig. 1 that increases with Z. Unlike the similar calculation for the BB-model in Fig. 7, the data here on average, tend to follow the SC-theoretical curve. A hypothetical 0.1 error of Z appears to bound the noisy data.

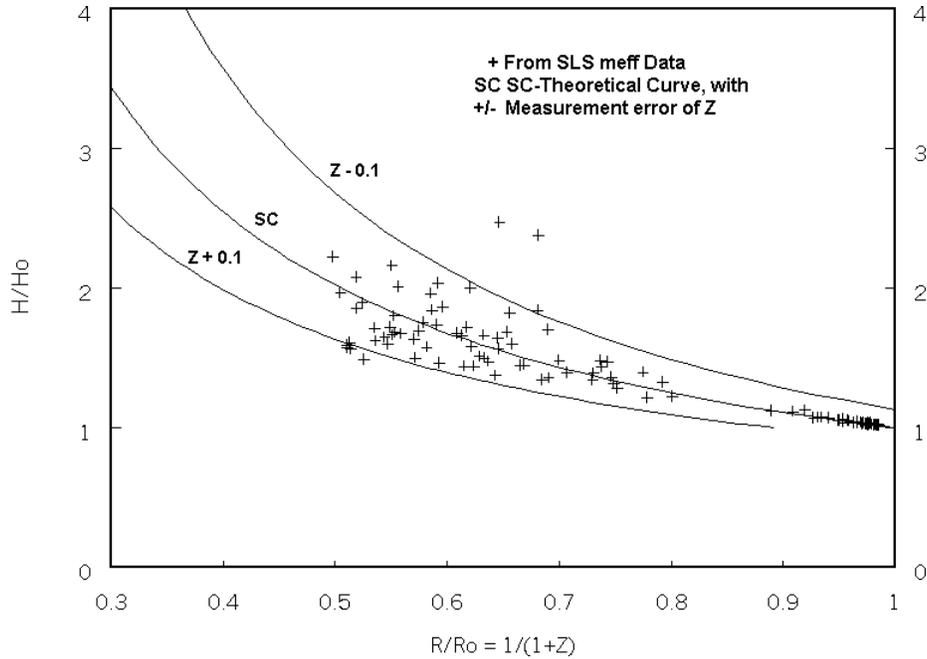

Fig. 9 The procedure for Fig. 8 was also performed on the more recent "Supernova Legacy Survey" data and the results are similar to Fig. 8.

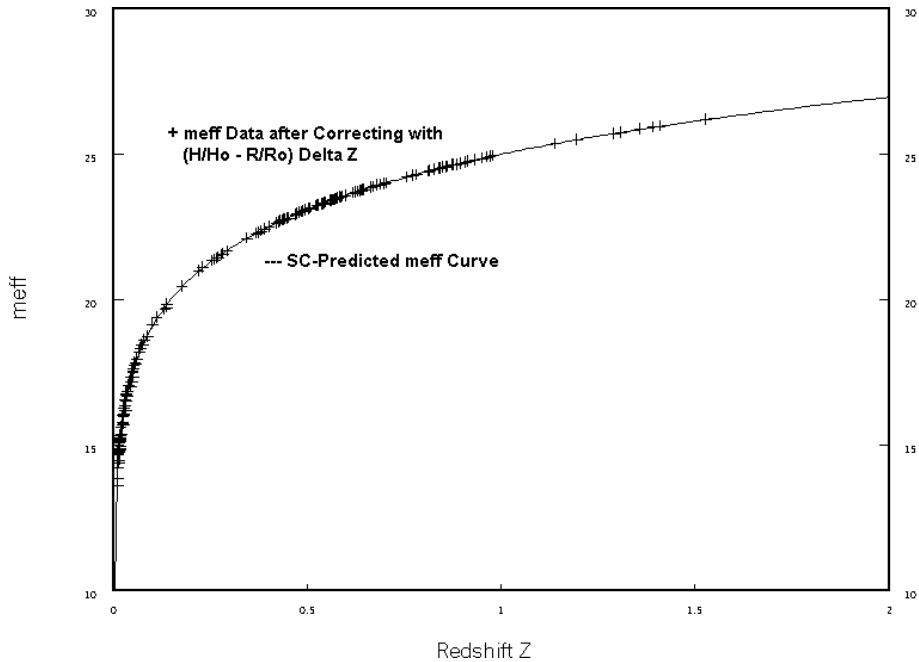

Fig. 10 The $(H/H_0 - R/R_0)\,\Delta Z$ corrections to fit theory in Fig. 8, were added to the raw data of Fig. 1 and greatly reduced the scatter of the data. Each data point now has a value of the Hubble parameter $H(Z)$ in agreement with theory. A similar agreement was obtained by adjusting $m_{eff}$.

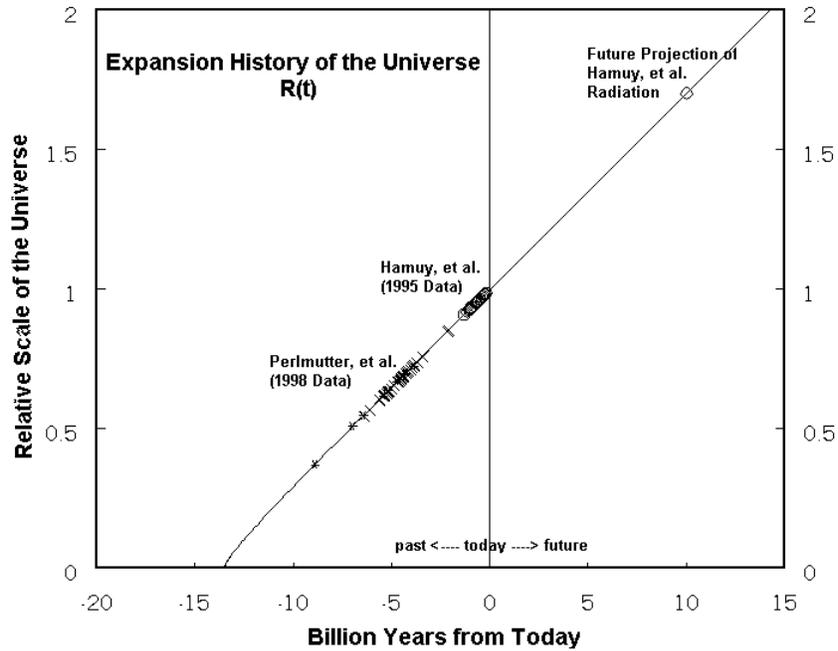

Fig. 11 The expansion history of our universe R to its present size and age $t_0$ =13.5 Gy. Consistency of the theory is shown by projection of the low-Z Hamuy, et al. data far into the future, $f = R_0'/R_0 = 1.70$ and $t_0 = 23.5$ Gy, where the theory there $R_0'$ correctly predicts the parameters at the present $R_0$ (see text).

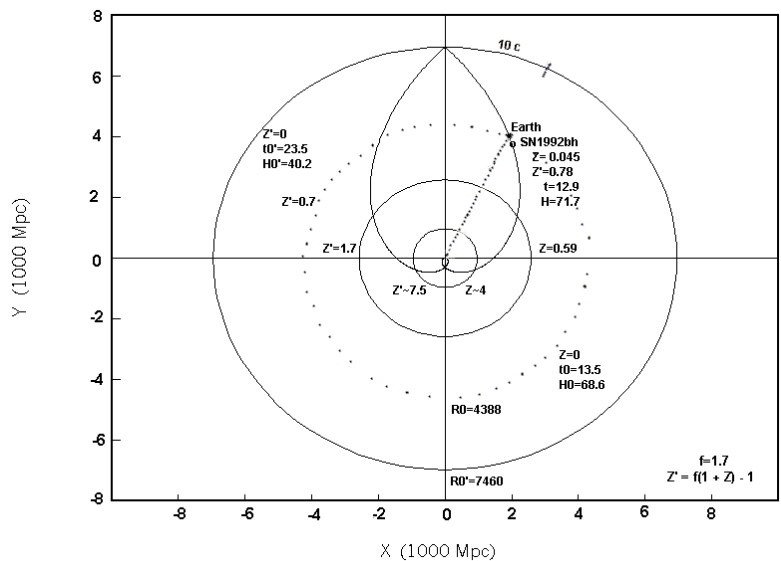

Fig. 12 In two of the 4-D coordinates, measurements of the Hamuy SN1992bh radiation are shown to the present $R_0$ and into the future to our universe $R_0'$ at a factor of 1.7 times its present size. The SC-theory predicts the parameters at $R_0'$ and from there predicts the present values and the same values of t and H at the source as were predicted from the present.

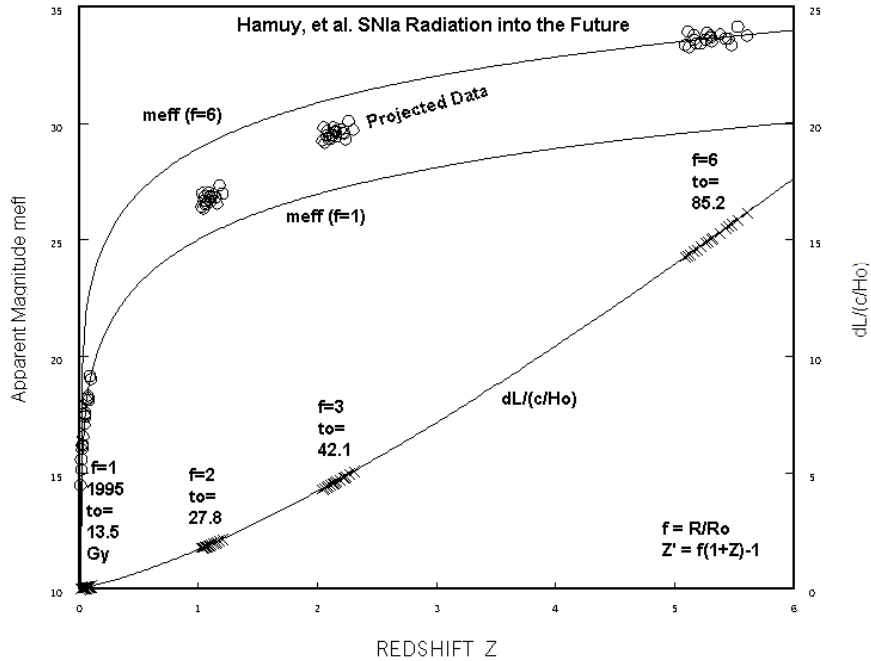

Fig. 13. Hamuy measurements are projected beyond instrument limitations of measurement. If the scale factor $R_0$ of our universe increases by a factor f, then the new $Z`$ of that radiation is $Z`=f(1-Z)-1$. The SC- luminosity distance moves up the $d_L/(c/H_0)$ universal curve while the effective luminosity $m_{eff}$ increases with decreasing flux.

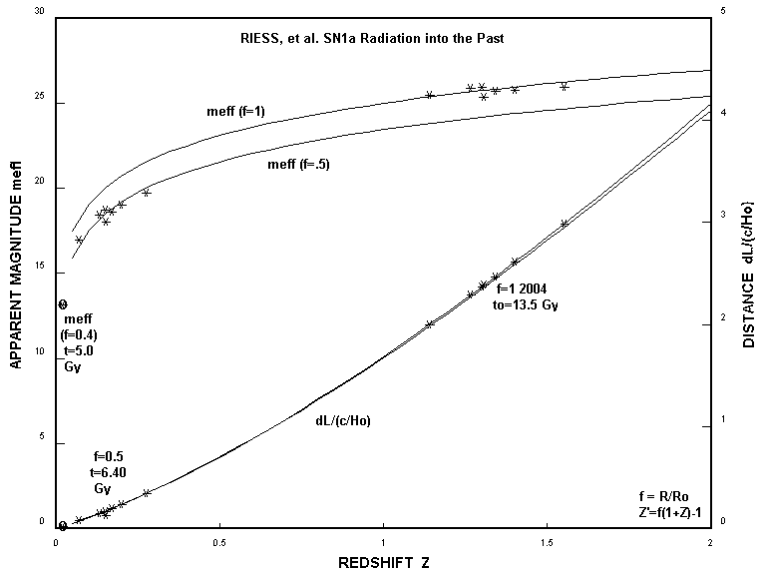

Fig. 14 Seven of the "Gold, High-Z" recent SNIa measurements of Riess, et al. [13] were predicted for past astronomer measurements at f=0.50 and t = 6.42 Gy. Again the same universal curve for $d_L/(c/H_0)$ is confirmed with $m_{eff}$ lower (energy flux higher) closer to the source. With f=0.40, and t=5.0 Gy, only the highest Z=1.551 SNIa appears because the other six have not yet exploded.